\documentclass[conference]{IEEEtran}
\usepackage{graphicx}
\usepackage{subcaption}
\usepackage{epsfig}
\usepackage{epstopdf}
\usepackage{color}
\usepackage{multirow}
\usepackage{booktabs}
\usepackage{amsmath,amsthm}
\usepackage{rotating}
\usepackage{algorithm}
\usepackage{algorithmic}
\usepackage{array}
\usepackage{listings}
\usepackage{animate}
\definecolor{dkgreen}{rgb}{0,0.6,0}
\definecolor{gray}{rgb}{0.5,0.5,0.5}
\definecolor{mauve}{rgb}{0.58,0,0.82}

\begin{document}
%
% paper title
% can use linebreaks \\ within to get better formatting as desired
\title{Distributed and Application-aware Task Scheduling in Edge-clouds}

\author{\IEEEauthorblockN
{Li Lin$^\dag$ \ Peng Li$^\ddag$ \ Jinbo Xiong$^\dag$  \ Mingwei Lin$^\dag$ \\}
\IEEEauthorblockA{\small
$^\dag$College of Mathematics and Informatics, Fujian Normal University, China \\
$^\ddag$School of Computer Science and Engineering, The University of Aizu, Japan \\
$^*$Corresponding author: Li Lin (llfjfz@163.com)}
}

\maketitle

\begin{abstract}
%\boldmath
Edge computing is an emerging technology which places computing at the edge of the network to provide an ultra-low latency. Computation offloading, a paradigm that migrates computing from mobile devices to remote servers, can now use the power of edge computing by offloading computation to cloudlets in edge-clouds. However, the task scheduling of computation offloading in edge-clouds faces a two-fold challenge. First, as cloudlets are geographically distributed, it is difficult for each cloudlet to perform load balancing without centralized control. Second, as tasks of computation offloading have a wide variety of types, to guarantee the user quality of experience (QoE) in terms of task types is challenging. In this paper, we present Petrel, a distributed and application-aware task scheduling framework for edge-clouds. Petrel implements a sample-based load balancing technology and further adopts adaptive scheduling policies according to task types. This application-aware scheduling not only provides QoE guarantee but also improves the overall scheduling performance. Trace-driven simulations show that Petrel achieves a significant improvement over existing scheduling strategies.

\end{abstract}

\begin{IEEEkeywords}
Edge computing, computation offloading, edge-clouds, task scheduling.
\end{IEEEkeywords}

% For peer review papers, you can put extra information on the cover
% page as needed:
% \ifCLASSOPTIONpeerreview
% \begin{center} \bfseries EDICS Category: 3-BBND \end{center}
% \fi
%
% For peerreview papers, this IEEEtran command inserts a page break and
% creates the second title. It will be ignored for other modes.

\section{Introduction}
\label{sec:introduction}

Edge computing is an emerging technology~\cite{Sat:EdgeComputing:definition}, which performs data analytics and storage close to the data source (i.e., mobile devices) to reduce the network latency. This computing paradigm has attracted great attention from both academic and industry. As a typical use case, computation offloading~\cite{MCCsurvey}, which migrates computing from mobile devices to the cloud, can now use the power of edge computing. The computation offloading over edge computing is also known as mobile edge computing~\cite{MEC:WhitePaper:5G}. By the deployment of edge-clouds, which are clusters of small servers (e.g., cloudlets~\cite{Cloudlet}) nearby mobile devices, the end user can benefit from the ultra-low latency and perform computation offloading in edge-clouds.

However, given a large number of end users asking for computation offloading services, the task scheduling in edge-clouds is a two-fold challenge. First, cloudlets in edge-clouds are geographically distributed, and task requests from end users dispersed to cloudlets are naturally unbalanced. Unlike the cloud, which usually employs monolithic schedulers that have centralized scheduling policies and complete control over all resources, cloudlets schedule tasks alone and concurrently. Without the global view of the overall cloudlets in edge-clouds, it is difficult for each cloudlet to perform load balancing strategies.

Secondly, mobile applications have a wide variety of types, and it is challenging to guarantee the user quality of experience (QoE) in terms of task types. Generally speaking, people are sensitive to the response latency when offloading latency-sensitive applications (e.g., augmented reality); however, they are more tolerant to the delay with latency-tolerant applications (e.g., deep neural net) but pay attention to if these applications would complete within specific latency bounds. Then, it puts forward a problem that how to schedule tasks efficiently in terms of task types, which is unsolved in the task scheduling for edge-clouds.

To solve the above challenges, in this paper, we propose Petrel, a distributed and application-aware task scheduling framework for edge-clouds. Unlike the monolithic schedulers in the cloud, Petrel is a lightweight scheduler deployed on each cloudlet running automatically. Petrel implements a simple \textit{sample-based} load balancing strategy, which employs the technology ``the power of two choices"~\cite{PowerOfTwo}. With this technology, for load balancing, Petrel randomly probes two cloudlets in edge-clouds and selects the cloudlet with the less load to place the task. This way has been proved to be effective under limited information~\cite{PowerOfTwoSurvey} and can also reduce the scheduling overhead significantly.

Furthermore, Petrel implements an application-aware scheduling algorithm, which adapts different scheduling policies in terms of task types. Specifically, for latency-sensitive tasks, Petrel uses a ``greedy" policy to find the cloudlet which has the minimum completion time for the tasks; whereas, for latency-tolerant tasks, Petrel adopts a policy of ``best effort" scheduling. In the ``best effort" scheduling, if there are idle resources on the cloudlet, Petrel then performs the task assignment; otherwise, it delays the task for a while but with the latency bound guarantee, which we call \textit{delay scheduling}. The application-aware scheduling achieves a significant performance improvement which is substantiated in our analysis and experiments.

In this paper, our contributions can be summarized as follows:
\begin{itemize}
  \item we propose Petrel, a distributed task scheduling framework for computation offloading in edge-clouds.
  \item Petrel uses a sample-based strategy for load balancing to reduce the scheduling overhead.
  \item Petrel implements an application-aware scheduling algorithm to improve the overall performance.
\end{itemize}

The rest of this paper is organized as follows. Section~\ref{sec:related:work} reviews some background and related work about the task scheduling in edge-clouds. Section~\ref{sec:system:architecture} presents an overview of the system architecture. Section~\ref{sec:scheduling:model} builds the model of task scheduling and objective. Section~\ref{sec:algorithm} introduces the distributed and application-aware task scheduling algorithm. Section~\ref{sec:evalution} evaluates Petrel with other existing strategies based on trace-driven simulations, followed by conclusions in Section~\ref{sec:conclusion}.

\section{Background and Related Work}
\label{sec:related:work}
In this section, we will review the background and related work.

\subsection{Edge Computing and Computation Offloading}
\label{sec:background:mec}
%Edge computing is now a compelling paradigm which extends the cloud computing to the edge of the network. The origins of edge computing can be traced to the development of the Internet of Things (IoT)~\cite{Bonomi:Fog:Computing} and 5G networks~\cite{MEC:WhitePaper:5G}, in which sensor data are ingested and analyzed at the edge for the low latency. However, as edge computing has various advantages, it enables a new breed of services and applications. As a typical scenario, computation offloading over edge computing, also known as mobile edge computing (MEC)~\cite{Patel:MEC:WhitePaper}, is an emerging computing paradigm that mobile devices migrate parts of their computation to the edge nodes in the vicinity. This paradigm offers a low end-to-end latency, which is critical for latency-sensitive applications.

Edge computing is now a compelling paradigm which extends the cloud computing to the edge of the network. The origins of edge computing can be traced to the development of the Internet of Things (IoT) and 5G networks~\cite{MEC:WhitePaper:5G}, in which sensor data are ingested and analyzed at the edge for the low latency. However, as edge computing has various advantages, it enables a new breed of services and applications. As a typical scenario, computation offloading over edge computing, also known as mobile edge computing (MEC), is an emerging computing paradigm that mobile devices migrate parts of their computation to the edge nodes in the vicinity. This paradigm offers a low end-to-end latency, which is critical for latency-sensitive applications.

%Computation offloading, known as mobile cloud computing (MCC)~\cite{MCCsurvey}, is a computing paradigm in which mobile devices migrate part of their computing to the cloud. In this way, computation offloading can reduce the energy consumption of mobile devices and speed up mobile applications. However, clouds are located far from mobile devices leading to a long communication delay, which would be disastrous for latency-sensitive applications, like gaming and augmented reality. Fortunately, edge computing is a promising way for computation offloading to solve the delay problem.

To utilize the power of edge computing, a large number of edge nodes are required to be deployed in the proximity to end users. These edge nodes are heterogeneous and different in the form factors. However, the most notable paradigm of the edge node is ``cloudlet"~\cite{Cloudlet}, a small data center nearby end users. Clusters of cloudlets, which are interconnected but without centralized nodes, become a kind of ``edge-clouds". Specifically, we use the word ``edge-cloud" to represent this infrastructure of cloudlets in this paper.

%Computation offloading supports a wide variety of mobile applications, and they are different in categories and the demand for resources~\cite{Silva2016:Offloading:Benchmark}. We classify these applications into two types: latency-sensitive and latency-tolerant. A latency-sensitive application requires a crisp response for the quick user interaction, such as augmented reality, gaming, face recognition, etc., and so the latency is a vital metric of the user quality of experience (QoE). In contrast, people are more tolerant of latency-tolerant applications. Generally, this kind of applications usually has intensive computing which is resource-demanding and time-consuming, like some scientific computing, deep neural net, antivirus, etc.

\subsection{Task Schedulers}

%Task scheduling is well-studied in the era of cloud computing~\cite{Delay:Scheduling}. Clusters in the cloud usually use monolithic schedulers, which employ a single centralized scheduling policy for all tasks. These schedulers have complete control over resources of all clusters. For example, YARN~\cite{YARN} is a typical monolithic scheduler, which provides a two-level resource scheduling framework in Hadoop. To optimize the scheduling concurrency, Schwarzkopf \textit{et al.} present the Omega~\cite{Omega}, which uses a shared state approach enabling each scheduler to fully access the entire clusters to improve the parallelism.
%
%However, with several orders of magnitude growing of the scale of clusters, the monolithic schedulers cannot satisfy the latency demand of task scheduling. Therefore, clusters are shifting to decentralized designs where multiple schedulers perform task scheduling concurrently. Sparrow~\cite{Sparrow} is one of the typical decentralized schedulers, which provides a distributed and low latency scheduling framework towards large-scale data analytics in Spark\footnote{http://spark.apache.org/}.
%To solve the problem of load balancing, Sparrow adapts "the power of two choices" technique~\cite{PowerOfTwo}, in which for each task the scheduler randomly queries the load of two servers and assigns the task to the least loaded of the two servers~\cite{PowerOfTwoSurvey}.

The research of task scheduling in edge-clouds is fresh. As cloudlets are geo-distributed, scheduling on cloudlets is naturally decentralized. Rashidi \textit{et al.}~\cite{Rashidi2017} have proposed a dynamic cloudlet selection policy, which uses the neuro-fuzzy inference system to place tasks. This policy is robust to the limited scheduling information. Shi \textit{et al.}~\cite{SHI2016} have presented an adaptive probabilistic scheduler, which can optimize the energy consumption of tasks with time-constrained. Zhao \textit{et al.}~\cite{Zhao2015:CooperativeScheduling} have introduced a cooperative scheduling mechanism over edge-clouds and the cloud. The above works propose sophisticated algorithms to place tasks on cloudlets across the edge-clouds but are unaware of the characteristic of task types, which has a vital impact on the task scheduling. By contrast, Petrel implements a lightweight load balancing strategy and further an application-aware task scheduling algorithm.

\section{System Architecture}
\label{sec:system:architecture}

\begin{figure}
\centering
\includegraphics[width=0.4\textwidth]{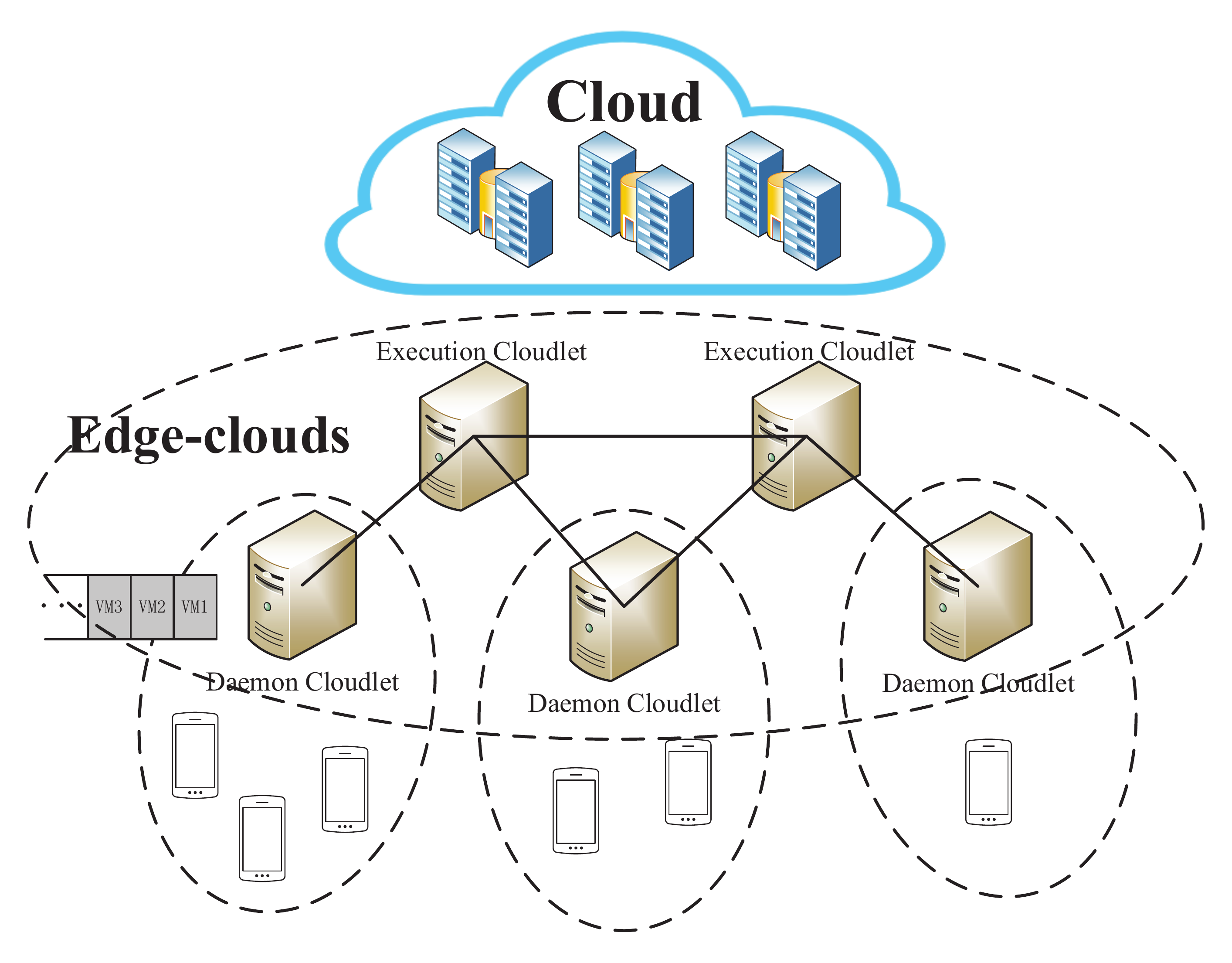}
\caption{System architecture}
\label{fig:architecture}
\vspace{-10pt}
\end{figure}

%A typical scenario of edge computing consist of multiple participants, including mobile devices, edge and cloud, which are associated with heterogeneous network. Figure~\ref{fig:architecture} illustrates the system architecture used in this paper. Mobile devices connect edge nodes with wireless network by APs (access point), and they are also able to directly access to cloud. It means that mobile devices can perform computation offloading to cloud or edge nodes. Edge nodes here refer to the facility of cloudlets~\cite{Cloudlet}, which are well connected to Internet and stay close to mobile devices. We propose the concept of "Daemon Edge", which acts as a best nominative edge responsible for computing proxy at an area of local network. A daemon edge responses to computing requests from mobile devices and dispatches their tasks on account of scheduling policies.
%
%(Service register, discovery, VM setup, how to scheduling)
%In the architecture, we suppose edge nodes connect with each other. It means that if there are N edge nodes, and the \emph{i}th node connect to the other \emph{N-1} nodes with different network RTT. We use the VM-based cloudlets~\cite{Justintime} as the computing platform of edge nodes. Every offloading task is assigned to a VM, and one time a VM can only serve one offloading task.

Fig.~\ref{fig:architecture} illustrates the architecture of Petrel. It is a typical multi-tiered architecture, including mobile devices, edge-clouds, and the cloud.
%Mobile devices, such as smartphones, tablets, etc., equipped with mobile applications, seek powerful remote servers for computation offloading.
Thanks to the deployment of edge-clouds over cloudlets which are well-connected with each other, mobile devices can migrate their computation to these cloudlets for low latency. We introduce the concept of ``daemon cloudlet", which has the lowest delay with mobile devices in the vicinity, and it plays the role of a computing proxy for the mobile devices. Ideally, an offloading task is served by its daemon cloudlet; however, too many tasks can cause a long queueing time affecting the performance, and so a daemon cloudlet would send tasks to other execution cloudlets for load balancing. Petrel copes with the problem that whether a task would be executed on a daemon cloudlet or other execution cloudlets and which execution cloudlet to choose. Notice that daemon cloudlets and execution cloudlets are relative, a cloudlet can be a daemon cloudlet for some mobile devices but an execution cloudlet for others.
%Finally, the cloud on the top of the architecture, offering a global view of edge-clouds, can help cloudlet discovery for daemon cloudlets and be candidate for offloading running. (Cloud is not used in the scheduling, and should clarify daemon cloudlets and executing cloudlets)

To support a standard execution environment of computation offloading, cloudlets in edge-clouds are built on virtual machines (VMs). VMs based resource provisioning~\cite{Justintime} is a widely-used technology to set up the executing environment for cloudlets quickly.
%VMs encapsulate the heterogeneity of mobile devices and cloudlets and create consistent environments for them.
When a computation offloading executed, a mobile device migrates its method execution to a VM associated with uploading data about the method. Then when the cloudlet finishes, it returns results to the mobile device. Every daemon cloudlet makes scheduling decisions automatically for task requests from mobile devices in the vicinity. The cloud in Petrel is backup for task executing but not a centralized scheduling node.

\section{Scheduling Model}
\label{sec:scheduling:model}

%调度模型，在daemon上，如果不行迁移到其他节点，云
In this section, we build a model for the task scheduling in edge-clouds, where the ``task" refers to a computation offloading from a mobile device to a remote server.
A task can be executed on mobile devices, cloudlets in edge-clouds or the cloud.
%Generally speaking, a task is first considered to be executed on the daemon cloudlet because of its lower latency. However, the performance of a daemon cloudlet degrades on account of too many task requests. As a result, the daemon cloudlet distributes tasks to other execution cloudlets for load balancing.
It is a multi-cloudlets scheduling problem, and we define the problem in the following.
%(Something about the decision algorithm???)

\subsection{Task Completion Time}
\label{sec:scheduling:model:ct}

%First, we define a task set of $J$, which contains task requests for end devices geo-distributed; task arrivals in $J$ are independent and identically distributed (i.e., i.i.d). For, each task $i$ in $J$, the task completion time on an end device, a cloudlet and the cloud is define in

The set $J$ contains tasks from mobile devices; the task arrivals are independent and identically distributed (i.e., i.i.d). For each task $i$ in $J$, the completion time of the task on the mobile device, the cloudlet, and the cloud are defined as:
\begin{equation}
\label{eq:ct:define}
T_{i} = \left\{
\begin{array} {l}
T_{i}^{mobile}, \text{the mobile device}; \\
T_{i}^{cloud}, \text{the cloud}; \\
T_{i}^{cloudlet}, \text{the cloudlet}.
\end{array}
\right.
\end{equation}
The completion time denotes the time span which a specific platform completes the task.

When a task runs on the mobile device, the completion time is equal to the task execution time $R_{i}^{mobile}$
\begin{equation}
\label{eq:ct:mobile}
T_{i}^{mobile}=R_{i}^{mobile},
\end{equation}
where $R_{i}^{mobile}$ depends on the hardware of the mobile device.

If a task is decided to be executed in the cloud, and its completion time is calculated by
\begin{equation}
\label{eq:ct:cloud}
T_{i}^{cloud}=R_{i}^{cloud}+\frac{D_{i}}{B_{cloud}}+RTT_{cloud}.
\end{equation}
$R_{i}^{cloud}$ is the task execution time in the cloud; $D_{i}$ indicates the data volume of uploading and downloading, and $B_{cloud}$ is the bandwidth; $RTT_{cloud}$ is the network delay. The part of $\frac{D_{i}}{B_{cloud}}+RTT_{cloud}$ denotes the communication time between the mobile device and the cloud. In general, the cloud is considered to have unlimited resource capacity, and so a task arriving in the cloud is immediately served with the time $R_{i}^{cloud}$.

The case of executing tasks on cloudlets is complicated. Cloudlets in edge-clouds can be seen as a network topology $G=(V,E)$, where the set $V$ denotes the connected cloudlets. In Petrel, a daemon cloudlet has the lowest latency with mobile devices in the vicinity. If a task is assigned to its daemon cloudlet $v_{d}\in V$, the completion time is
\begin{equation}
\label{eq:ct:daemon:cloudlet}
T_{i}^{cloudlet}=R_{i}^{v_{d}}+W_{i}^{v_{d}}+\frac{D_{i}}{B_{v_{d}}}+RTT_{v_{d}}
\end{equation}

Unlike the cloud, cloudlets have limited resource, and the task should contend for running including the waiting time $W_{i}^{v_{d}}$. As a result, the completion time contains the task execution time $R_{i}^{v_{d}}$ on the $v_{d}\in V$, the waiting time $W_{i}^{v_{d}}$, and the communication time $\frac{D_{i}}{B_{v_{d}}}+RTT_{v_{d}}$. Generally, cloudlets have larger bandwidth $B_{v_{d}}$ and lower delay $RTT_{v_{d}}$ than the cloud, and so the smaller of the communication time.
%The benefit of fast data communication can overtake the additional waiting time if a well-designed scheduling algorithm is used.
In particular, for load balancing, a daemon cloudlet would send the task to other execution cloudlets, for example, the cloudlet $v_{e}\in V$. If the task $i$ is assigned to the cloudlet $v_{e}$, and its completion time is calculated by:
\begin{equation}
\label{eq:ct:executor:cloudlet}
T_{i}^{cloudlet^{'}}=R_{i}^{v_{e}}+W_{i}^{v_{e}}+\frac{D_{i}}{B_{v_{e}}}+RTT_{v_{d}}+RTT_{v_{e}}
\end{equation}

In (\ref{eq:ct:executor:cloudlet}), it has an additional $RTT_{v_{e}}$ comparing to (\ref{eq:ct:daemon:cloudlet}). As in Petrel, the mobile device first connects to its daemon cloudlet which decides how to place tasks; if the task is finally assigned to other execution cloudlets, e.g., $v_{e}$ in this example, the daemon cloudlet will redirect the task to $v_{e}$ with the additional delay $RTT_{v_{e}}$.

\subsection{The Scheduling Objective}
\label{sec:scheduling:model:objective}

To formalize the completion time on different platforms, we define an allocation vector $A=(\alpha_{1},\alpha_{2},\alpha_{3})$ to indicate which platform a task is assigned to. In the vector $A$, $\alpha_{1}$ denotes whether the task is executed on a mobile device with $\alpha_{2}$ for the cloud and $\alpha_{3}$ for the cloudlet; the value 1 represents \textit{yes}, otherwise 0. For example, $A=(0,0,1)$ indicates that the task is executed on the cloudlet. Further, we use the task speedup as the metric indicating the benefit of the task executed on remote servers
\begin{equation}
\label{eq:metric}
Sp_{i}=\frac{R_{i}^{mobile}}{AT_{i}}.
\end{equation}
Then, for all tasks in $J$, the objective of task scheduling is to maximize the following:
\begin{equation}
\label{eq:objective}
\overline{Sp_{i}}=(\sum_{i\in J}\frac{R_{i}^{mobile}}{AT_{i}})/N,
\end{equation}
i.e., the average task speedup, where $N$ indicates the number of task. Specifically, for each task $i$, the $Sp_{i}$ should larger than 1, which means the computation offloading should improve its performance on the mobile device.
%(something about latency bound)

\section{Distributed and Application-aware Task Scheduling}
\label{sec:algorithm}

\begin{figure}
\centering
\includegraphics[width=0.35\textwidth]{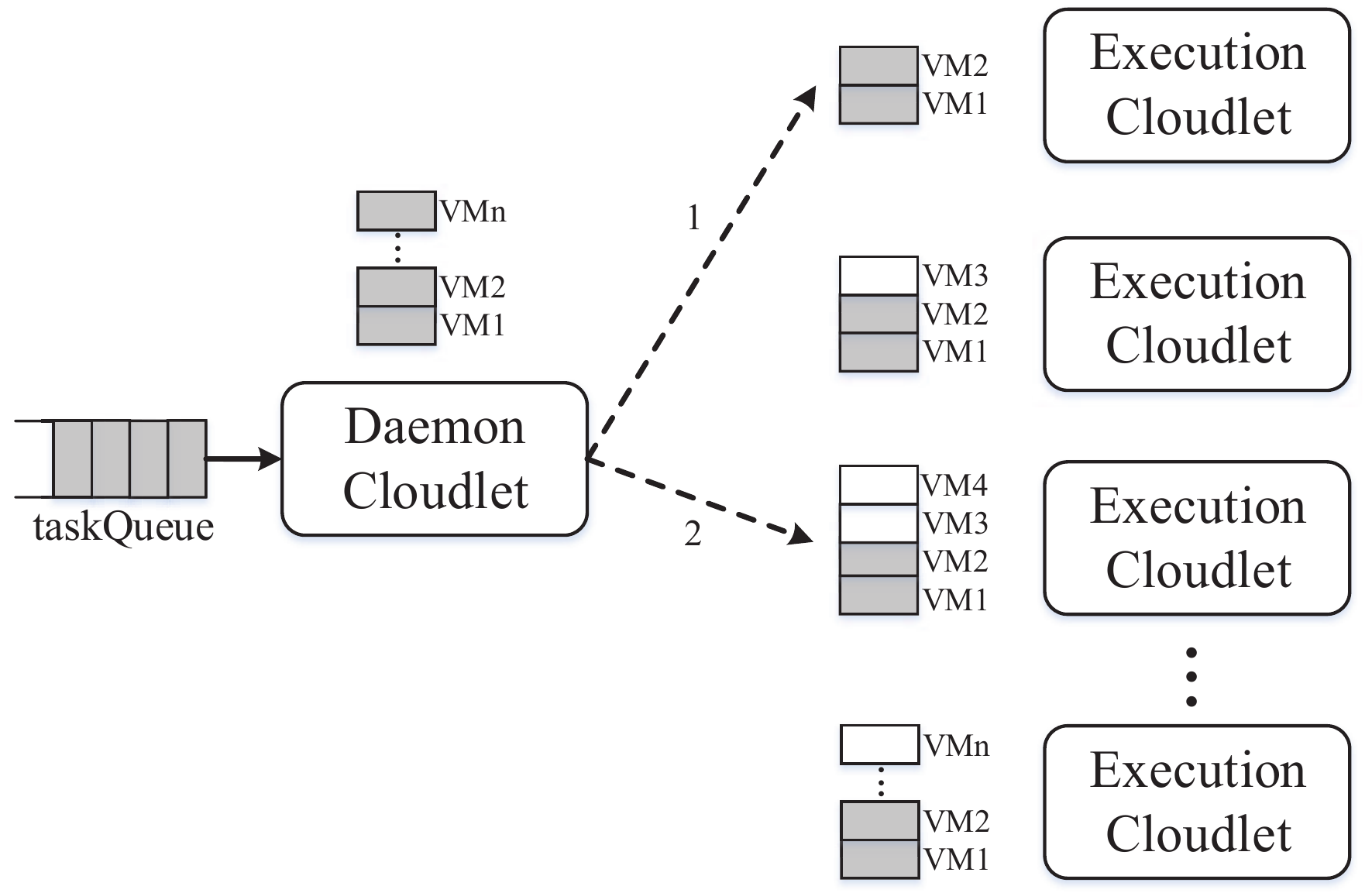}
\caption{An illustration of task scheduling in Petrel. If the daemon cloudlet has no idle VMs, it will probe two other execution cloudlets and pick up the one with the less load.}
\label{fig:scheduling:demo}
\vspace{-10pt}
\end{figure}

To solve the challenges of the task scheduling in edge-clouds, in this section, we propose a distributed and application-aware algorithm (DAA).

Fig.~\ref{fig:scheduling:demo} illustrates the process of task scheduling in Petrel. Tasks are served using the principle of first come, first served (FCFS) based on VMs as one task at a time for each VM. Once a task is scheduled, Petrel first finds if there are idle VMs on the daemon cloudlet, if so, then the daemon cloudlet will execute the task; otherwise, Petrel will probe other execution cloudlets. As shown in Fig.~\ref{fig:scheduling:demo}, when the task at the head of \textit{taskQueue} is scheduled, the daemon cloudlet finds itself has no idle VMs at the time; then it sends probe \textit{1} and \textit{2} and finds the probe \textit{2} with the less load to assign the task.
%(VM numbers are the power)

The details of task scheduling are shown in Algorithm~\ref{alg:daa}. For each $task_{i}$ in \textit{taskQueue}, if the daemon cloudlet has idle VMs, it will serve the task immediately (in line 2); otherwise, it performs the load balancing policy (in line 5) and adopts different assignment strategies in terms of task types (in lines 6 to 25).

\begin{algorithm}[t]
\renewcommand{\algorithmicrequire}{\textbf{Initialization:}}

\caption{Distributed and Application-aware Scheduling}
\label{alg:daa}
\begin{algorithmic}[1]
\REQUIRE for each $task_{i}$ arrives at its daemon cloudlet $cloudlet_{d}$, we insert $task_{i}$ into the queue $taskQueue$;
$Type_{sen}$ denotes the set of latency-sensitive tasks; $bound_{i}$ denotes the latency bound of $task_{i}$
%\ENSURE
\IF {$cloudlet_{d}$ has idle VMs}
    \STATE assign $task_{i}$ to $cloudlet_{d}$;
\ELSE
    \STATE calculate the expected completion time $t_{i}^{d}$ on $cloudlet_{d}$;
    \STATE randomly probe two cloudlets in edge-clouds, and pick up the one, $cloudlet_{e}$, with the least loaded of the two cloudlets;
    \IF {$task_{i} \in Type_{sen}$}
        \STATE calculate the expected completion time $t_{i}^{e}$ on $cloudlet_{e}$;
        \IF {$t_{i}^{e} < t_{i}^{d}$}
            \STATE assign $task_{i}$ to $cloudlet_{e}$;
        \ELSE
            \STATE assign $task_{i}$ to $cloudlet_{d}$;
        \ENDIF
    \ELSE
        \STATE // $task_{i}$ is a latency-tolerant task
        \IF {$cloudlet_{e}$ has idle VMs}
            \STATE assign $task_{i}$ to $cloudlet_{e}$;
        \ELSE
            \STATE calculate the expected completion time $t_{i}^{d'}$ on $cloudlet_{d}$ for a delay of time $D$;
            \IF {$t_{i}^{d'} \ge bound_{i}$  }
                \STATE assign $task_{i}$ to $cloudlet_{d}$;
            \ELSE
                \STATE delay $task_{i}$ for time $D$
            \ENDIF
        \ENDIF
    \ENDIF
\ENDIF
\end{algorithmic}
\end{algorithm}

The ``load" in Algorithm~\ref{alg:daa} indicates the task queue length with respect to the task expected completion time on a cloudlet. Specifically, the task expected completion time means the wall-clock time at which a cloudlet completes a task. In DAA, what we say the least loaded of two cloudlets means if a task is tentatively assigned to the two cloudlets, the one providing the earlier expected completion time has the less load, as the $cloudlet_{e}$ in the algorithm. We maintain a \textit{PriorityQueue} to store the ready time for all VMs on a cloudlet, and so the head of the \textit{PriorityQueue} is the earliest ready time for a task. Then, the task expected completion time is calculated based on the \textit{PriorityQueue}.
%and whenever a task is assigned to a cloudlet on a specific VM, the \textit{PriorityQueue} will be reset, as in lines 2, 9, 11, 16, and 20.

%As tasks executed based on VMs, for each cloudlet, we maintain a PriorityQueue, which holds items that indicate the available time for all VMs. Therefore, the head of the PriorityQueue is always the first available VM which we will assign a task to. Moreover, if a task is assigned to a cloudlet, its PriorityQueue will be resorted, as in line 2, 9, 11, 16, and 20. Furthermore, to calculate the expected completion time of a task is also based on the PriorityQueue, as in line 4, 7, and 18.

%If the daemon cloudlet has no , the load balancing technology--"the power of two choices" will be used, shown in line 5. Specially, "the load" does not mean the queue length but the expected completion time of a task. In other words, if $cloudlet_{m}$ has a less load, a task assigned to it can have an earlier expected completion time. Meanwhile, we adopt two different strategies in terms of task type.

If there are no idle VMs on the daemon cloudlet, then the task scheduling strategies are task type dependent. If $task_{i}$ is a latency-sensitive task, we compare its expected completion time on $cloudlet_{e}$ with that on the daemon cloudlet $cloudlet_{d}$ and assign the task to the faster one, as in lines 7 to 12. This comparison is necessary, as a task migrating to an execution cloudlet will cause an additional network delay shown in (\ref{eq:ct:executor:cloudlet}). As a result, this comparison guarantees the optimal latency of task executing. On the other hand, if $task_{i}$ is a latency-tolerant task, we first check if $cloudlet_{e}$ has idle VMs; if so, we perform the assignment; otherwise, we tentatively delay the task scheduling and calculate the expected completion time $t_{i}^{d'}$ in line 18. The $t_{i}^{d'}$ should satisfy the latency bound of $task_{i}$; if not, it will be assigned to $cloudlet_{d}$ with the expected completion time $t_{i}^{d}$. This process is what we call the ``best effort" scheduling with the \textit{delay scheduling} if there are no idle VMs.

\section{Evaluation}
\label{sec:evalution}
In this section, we evaluate the performance of Petrel based on trace-driven simulations. First, we introduce benchmarks used in the simulations. Then, the experimental methodology is provided.
%Finally, we show the results and offer a comprehensive analysis.

\begin{figure}
\centering
\includegraphics[width=0.35\textwidth]{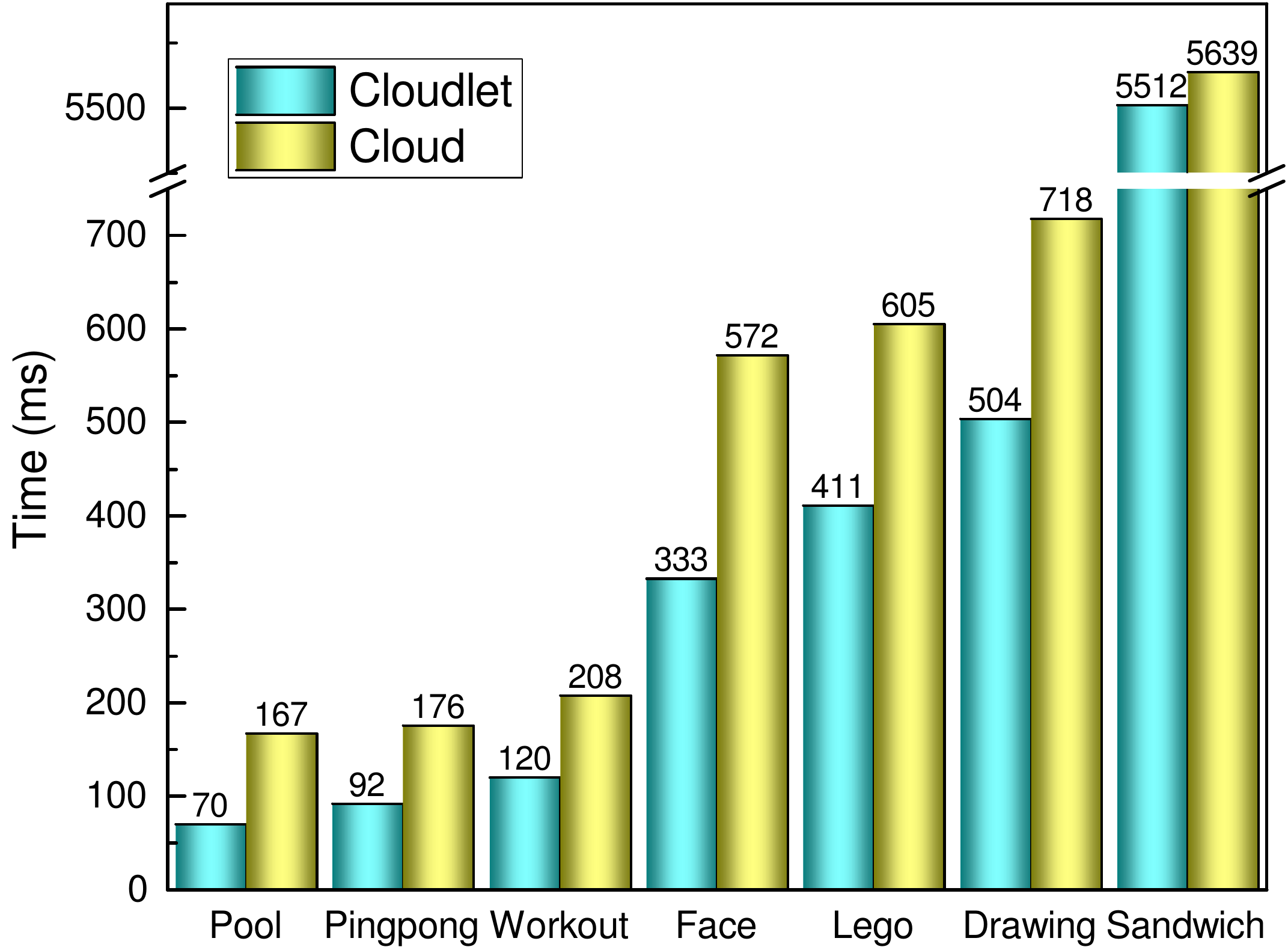}
\caption{The mean latency of benchmarks offloading to the cloudlet and the cloud}
\label{fig:benchmark:latency}
\vspace{-10pt}
\end{figure}

\subsection{Benchmarks}
Applications in computation offloading have various types and QoE demand~\cite{Silva2016:Offloading:Benchmark}, and we classify these applications into two categories: latency-sensitive applications and latency-tolerant applications. Then, we use the applications listed in the paper~\cite{Chen:2017:Edge:Benchmark} as benchmarks, which are type of cognitive assistance applications~\cite{Gabriel}.
%and the paper provides an empirical study of application latency on cloudlets and the cloud.
%Table~\ref{tab:benchmarks} illustrates the details of the benchmarks, and these benchmarks are the type of cognitive assistance applications~\cite{Gabriel}.
Fig.~\ref{fig:benchmark:latency} shows the mean latency when these benchmarks are offloaded to the cloudlet and the cloud, source from~\cite{Chen:2017:Edge:Benchmark}.
%As can be seen, the benefit of the network proximity of cloudlets is obvious that cloudlets have a lower RTT than the cloud, as the analysis in Section~\ref{sec:scheduling:model:ct} refer to (\ref{eq:ct:cloud}) and (\ref{eq:ct:daemon:cloudlet}).
Notice that the latency on the cloudlet in this figure is considered as the \textit{service time} in the following simulations; the service time indicates the time span if a task runs alone on the cloudlet. By further analysis, we find that the \textit{Sandwich} uses a deep neural net approach which is compute-intensive and time-consuming, and people are more tolerant of the application; therefore, it can be seen as a latency-tolerant application. However, the other applications, employing some simple AI algorithms, need a tight interaction with users, and so they are latency-sensitive applications.

\subsection{Methodology}
To conduct the simulations, we build a trace-driven data set. The data set contains 200 offloading tasks; each task is randomly selected from benchmarks in Fig.~\ref{fig:benchmark:latency}.
%and the service time of tasks are illustrated in Fig.~\ref{fig:benchmark:latency}.
There are 10 cloudlets in our edge-cloud, and each cloudlet has the number of virtual machine range from 1 to 10. Task arrivals on their daemon cloudlets are modeled as a Poisson process with the arrival rate $\lambda$. We set two different arrival rates $\lambda$=1 and $\lambda$=2 in our simulations, which means the mean time intervals between two tasks are 1 and 0.5 unit time respectively.

The latency illustrated in Fig.~\ref{fig:benchmark:latency} can be seen as the time span when a task is offloaded from a mobile device to its daemon cloudlet, which is considered to have the lowest latency with mobile devices in the vicinity, and the RTT is usually smaller than 10ms. However, if a daemon cloudlet decides to send tasks to other execution cloudlets, it will cause an additional RTT from mobile devices to execution cloudlets according to (\ref{eq:ct:executor:cloudlet}), and this additional RTT usually ranges from 50-70ms~\cite{Agarwal:Vision:SmallCells}. We consider these RTTs in the construction of our data set.

\subsubsection{Comparing Algorithms}
For a comprehensive evaluation, we compare DAA with the following scheduling algorithms:
\begin{itemize}
  \item DaemonCloudletOnly: all tasks are executed only on daemon cloudlets with the rule of FCFS. There is no load balancing strategy.
  \item RoundRobin: a simple load balancing algorithm. Tasks are distributed to cloudlets on the edge-cloud in a round robin way. This algorithm treats each cloudlet equally, without considering the benefit of daemon cloudlets.
  \item GreedyScheduler: a greedy scheduling algorithm. For each task, GreedyScheduler always finds an optimal cloudlet which has the minimum completion time to serve the task.
  %Therefore, to make a decision, a daemon cloudlet should query every cloudlet on the edge-cloud for the load information, and so it incurs enormous scheduling overhead.
  \item TwoChoices: employing the ``the power of two choices" strategy. In TwoChoices, for each task, it randomly probes two cloudlets and selects the one has the less load.
\end{itemize}
Our proposed DAA in Algorithm~\ref{alg:daa}, is a distributed scheduling algorithm, employing a sample-based load balancing technology and further an application-aware scheduling strategy.

\subsection{Simulation Results}
In this section, we evaluate the DAA algorithm in two metrics: the average weighted turnaround time (AWT) and the makespan for cloudlets.

\subsubsection{Average Weighted Turnaround Time}

%(Changed to completion time)
The average weighted turnaround time is defined as:
\begin{equation}
\label{eq:weighted:turnaround}
AWT=(\sum_{i}^{N}{\frac{T_{i}^{turnaround}}{T_{i}^{service}}})/{N}.
\end{equation}
$T_{i}^{service}$ is the \textit{service time} as a task is served by a cloudlet alone; $T_{i}^{turnaround}$ denotes the turnaround time of a task, i.e., the \textit{completion time} we define in Section~\ref{sec:scheduling:model:ct}; $N$ is the number of tasks. Reviewing (\ref{eq:objective}) in Section~\ref{sec:scheduling:model:objective}, the scheduling objective is to maximize the overall task speedups. In other words, by comparing (\ref{eq:objective}) and (\ref{eq:weighted:turnaround}), it can be seen that \textit{the higher the average task speedup, the lower the average weighted turnaround time}.

\begin{figure}[tb]
	\centering
	\shortstack{
		\includegraphics[width=0.3\textwidth]{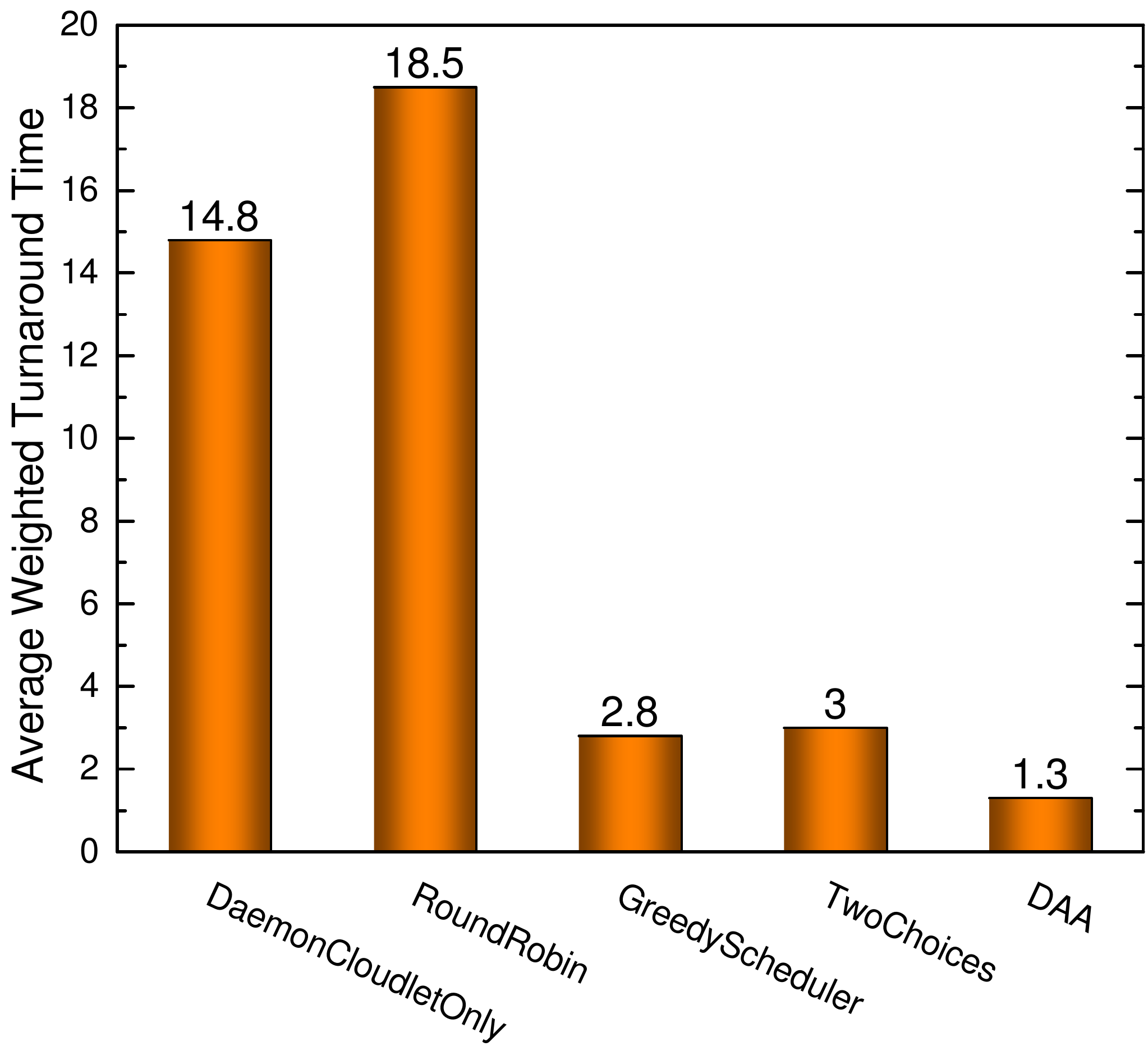}\\
		{(a) $\lambda=1$}
	}\quad
	\shortstack{
		\includegraphics[width=0.3\textwidth]{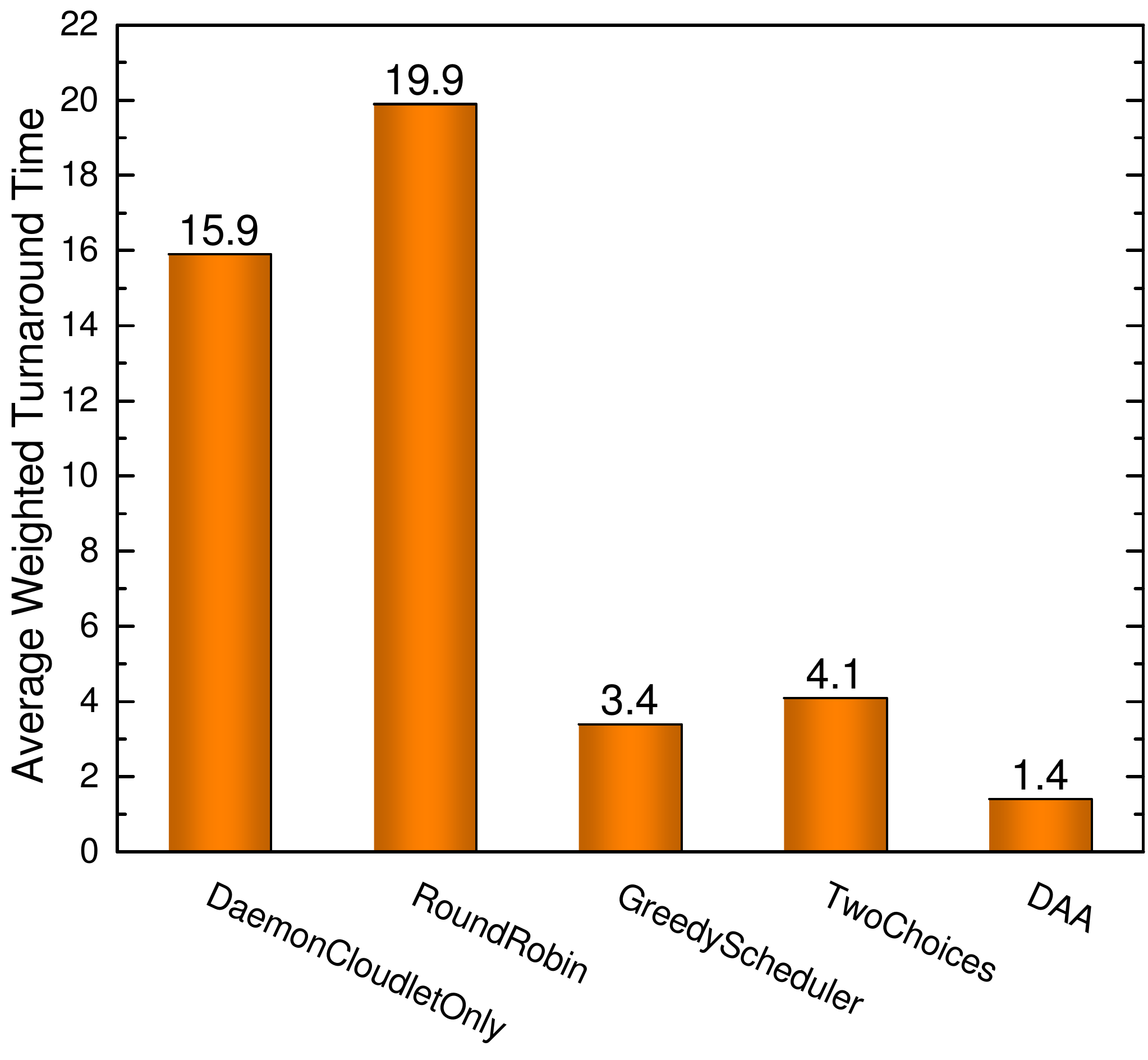}\\
		{(b) $\lambda=2$}
	}
	\caption{The average weighted turnaround time\label{fig:evaluation:awt}}
\vspace{-10pt}
\end{figure}

Fig.~\ref{fig:evaluation:awt} shows the comparison of these algorithms in terms of the average weighted turnaround time. As can be seen, DaemonCloudletOnly and RoundRobin have the worst performance. Specifically, DaemonCloudletOnly restricts all tasks to be executed only on daemon cloudlets and employs no load balancing strategies, leading to poor performance on the less powerful cloudlets. Although RoundRobin uses a simple Round Robin load balancing, this strategy treats every cloudlet equally without considering the benefit of daemon cloudlets and the more powerful cloudlets. GreedyScheduler achieves a significant performance improvement, as it always assigns tasks to the cloudlet that has the minimum completion time. However, it incurs enormous scheduling overhead across cloudlets on the edge-cloud. On the contrary, the TwoChoices algorithm only samples two cloudlets when performing load balancing, but it gets similar performance with GreedyScheduler.

It can be seen that DAA has better performance than the other algorithms, and the gap is increasing with a more frequent task arrival ($\lambda=2$). DAA adopts different strategies according to the type of tasks, as ``greedy" for latency-sensitive tasks and ``best effort" for latency-tolerant tasks. These strategies can conquer the drawback by always placing tasks greedily without considering the type of the task, which can cause the starvation of short tasks (latency-sensitive tasks) because of the long tasks (latency-tolerant tasks) executing, just as GreedyScheduler does. Moreover, DAA gets the task assignments with a lightweight load balancing strategy like TwoChoices does. Furthermore, we evaluate the performance when all tasks are executed on the cloud, and the average weighted turnaround time is 1.6, which is also worse than DAA. It appears that task executed on cloudlets can achieve better performance than on the cloud if the right scheduling algorithms are adopted. Meanwhile, all user tasks executed on the public cloud seems unpractical in terms of resource consumption.

\subsubsection{Makespan}

\begin{figure}[tb]
	\centering
	\shortstack{
		\includegraphics[width=0.3\textwidth]{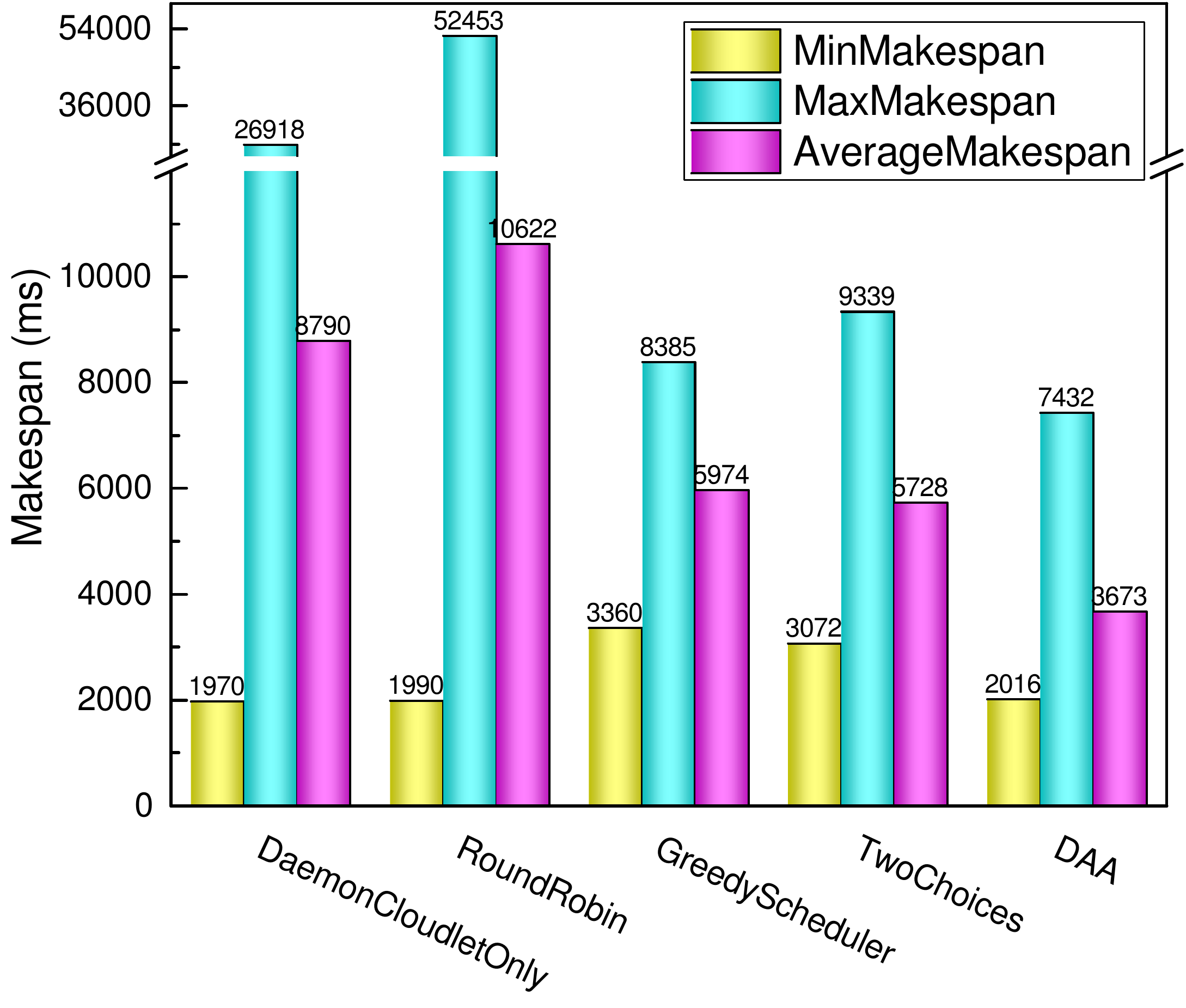}\\
		{(a) $\lambda=1$}
	}\quad
	\shortstack{
		\includegraphics[width=0.3\textwidth]{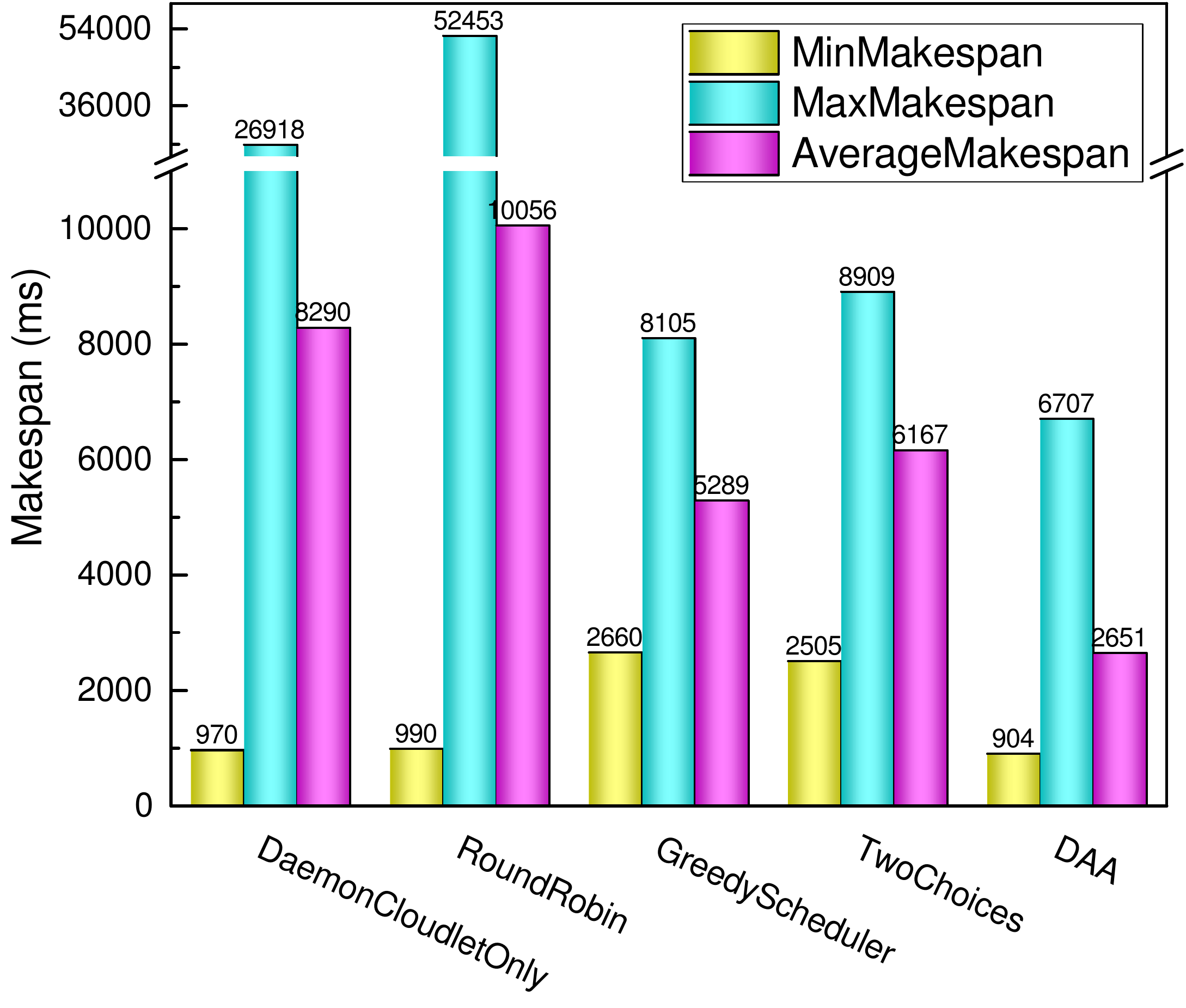}\\
		{(b) $\lambda=2$}
	}
	\caption{The minimum, maximum, and average makespan of cloudlets\label{fig:evaluation:makespan}}
\vspace{-10pt}
\end{figure}

Makespan is defined as the time when a cloudlet completes the last task; in other words, it is a metric of the throughput of the cloudlet~\cite{Maheswaran:Scheduling:algorithms}. Scheduling algorithms strive for the goal of minimizing the makespan and maximizing the throughput. Fig.~\ref{fig:evaluation:makespan} depicts the minimum (MinMakespan), the maximum (MaxMakespan), and the average makespan (AverageMakespan) of cloudlets. The average makespan indicates the mean value of all cloudlets in our edge-cloud. It can be seen that DaemonCloudlet and RoundRobin have the worst maximum and average makespan, as they perform no or inefficient load balancing strategies, which incurs so obviously unbalanced load among cloudlets. GreedyScheduler and TwoChoices improve the overall average makespan by distributing tasks efficiently, but sacrifice some cloudlets which have the lower load, as can be seen in the minimum makespan. Finally, the DAA algorithm achieves the best overall performance.

%\begin{figure}[tb]
%	\centering
%	\shortstack{
%		\includegraphics[width=0.4\textwidth]{figs/CompleteRate1.eps}\\
%		{(a) $\lambda=1$}
%	}\quad
%	\shortstack{
%		\includegraphics[width=0.4\textwidth]{figs/CompleteRateHalf.eps}\\
%		{(b) $\lambda=2$}
%	}
%	\caption{The task completion ratio\label{fig:evaluation:completerate}}
%
%\end{figure}

%\subsubsection{Completion Ratio}
%Each mobile application has a latency bound in terms of user experience, i.e., the maximum interactive latency that users can tolerate. Chen \textit{et al.} measure the latency bounds of benchmarks in~\cite{Chen:2017:Edge:Benchmark}; for example, the bound range for PingPong is 150-230ms. In their measurements, tasks can finish within the latency bounds by executing alone on the cloudlet. However, if many tasks contend for running, tasks may not satisfy the bound constraint on account of the waiting time.
%
%
%
%We define the task completion ratio as the ratio of the number of tasks which finish within latency bounds ($N_{\le{bound}}$) to the number of overall tasks ($N$):
%\begin{equation}
%\label{eq:chap3:completion:ratio}
%Completion~Ratio=N_{\le{bound}}/N.
%\end{equation}
%Fig.~\ref{fig:evaluation:completerate} shows the results of completion ratio. The results highly coincide with the above two metrics. First, DaemonCloudlet and RoundRobin have the worst performance. Then, GreedyScheduler and TwoChoices make some improvement, but not enough. Finally, DAA achieves a significant improvement; even if we double the task arrival rate, i.e., $\lambda=2$, DAA can still get a completion ratio up to 90\%.

\section{Conclusion}
\label{sec:conclusion}
In this paper, we present Petrel, a distributed and application-aware task scheduling framework for edge-clouds. Petrel implements a sample-based load balancing for cloudlets in edge-clouds, which is simple but efficient and can reduce the scheduling overhead sharply. Furthermore, Petrel adopts different scheduling policies in terms of task types, as the ``greedy" policy for latency-sensitive tasks but the ``best effort" service for latency-tolerant tasks. The results of trace-driven simulations show that our proposed scheduling strategies achieve significant improvements over the existing scheduling algorithms.

\section{Acknowledgments}
This work is supported by National Natural Science Foundation of China under grant No. 61502103, 61872088, and 61872086, and Natural Science Foundation of Fujian Province under grant No. 2017J01737.

\bibliographystyle{IEEEtran}
\bibliography{scheduling}

% that's all folks
\end{document}